\documentclass[secnumarabic, showpacs, amssymb, amsmath, nofootinbib,tightenlines,
nobibnotes, aps,prl]{revtex4}

\usepackage{amsmath,amssymb}
\usepackage{graphicx}

\begin{document}

\title{Aggregates relaxation in a jamming colloidal suspension after shear cessation}
\author{Francesca Ianni$^{1,2}$}
\email{francesca.ianni@phys.uniroma1.it}
\author{David Lasne$^{1,3}$}
\author{R{\'e}gis Sarcia$^1$}
\author{Pascal H{\'e}braud$^{1,4}$}
 \affiliation{$^1$ P.P.M.D. UMR $7615$ ESPCI $10$ rue Vauquelin $75231$ Paris Cedex 05, France\\
$^2$ SOFT-INFM-CNR, c/o Universit{\`a} di Roma ''La Sapienza'', I-00185, Roma, Italy\\
$^3$ C.P.M.O.H. UMR $5798$ 351 cours de la Lib{\'e}ration 33405
Talence, France\\
$^4$ IPCMS UMR $7504$ 23 rue du Loess 67034 Strasbourg Cedex 02,
France}

\begin{abstract}
The reversible aggregates formation in a shear thickening,
concentrated colloidal suspension is investigated through speckle
visibility spectroscopy, a dynamic light scattering technique
recently introduced  [P.K.~Dixon and D.J.~Durian, Phys. Rev. Lett.
90, 184302 (2003)]. Formation of particles aggregates is observed
in the jamming regime, and their relaxation after shear cessation
is monitored as a function of the applied shear stress. The
aggregates relaxation time increases when a larger stress is
applied. Several phenomena have been proposed to interpret this
behavior: an increase of the aggregates size and volume fraction,
or a closer packing of the particles in the aggregates.
\end{abstract}
\maketitle
\section{Introduction}
Concentrated colloidal suspensions exhibit complex rheological
behavior. At low stress, their viscosity decreases with increasing
stress, whereas it increases when the applied stress exceeds a
critical value. The increase of the viscosity may even lead to
cessation of the flow~: the suspension jams~\cite{lootens}. The
first phenomena, called shear thinning, has been extensively
studied and is associated with the advent of a long range order
between the particles, which align along the flow direction. On
the contrary, the particle microstructure responsible for the
shear-thickening phenomena at high stress is still not completely
understood. The mechanism responsible for shear thickening has
been studied numerically and theoretically in the simple shear
geometry. In that case, the stress tensor exhibits a compression
along an axis oriented at $3 \pi/4$ from the flow direction in the
flow-gradient plane. When the compressive force along this axis
overcomes repulsive inter-particles forces~\cite{wagner} (of
brownian, steric or electrostatic origin), anisotropic aggregates
of particles, oriented along the compression axis, form. These
aggregates induce a rapid increase of the
viscosity~\cite{melrose2,raiski,brady}. This has the further
effect of a sign reversal in the first normal stress difference,
which assumes negative values in strong shear
flows~\cite{zarraga}. The aggregate formation is
reversible~\cite{wagner2} and, according to simple model,
aggregates may span over the entire gap of the system~\cite{loan},
thus leading to flow instability~\cite{lootens}. The size
distribution of these shear-induced aggregates has been studied
in~\cite{loan} for a two dimensional simple driven diffusive model
and an hysteresis is found in the evolution of the average
aggregate size as a function of the shear rate. Numerical
simulations of concentrated hard spheres under
shear~\cite{melrose1}, taking hydrodynamic interactions into
account, show that the probability of having a percolating
aggregate with a given inter-particles spacing saturates when the
applied stress increases.
Moreover, the inter-particles spacing inside an aggregate decreases when the applied stress increases .\\
Experimental study of the shear induced aggregate formation is a
challenging issue. The first measurements investigating the
particle structure of a system in the shear thickening regime were
conducted through small angle neutron scattering~\cite{laun} and
proved the existence of a short range order and the absence of
long range order. In this kind of experiments \cite{watanabe}, the
quiescent scattering profile was recovered after shear cessation,
showing that the structure formed under shear is not of permanent
nature. On a mechanical point of view, the anisotropy of the
particles pair distribution function results in non-zero values of
the first and second normal stress differences. Nevertheless,
measurements of normal stress differences are extremely difficult
due to their very small value~\cite{gadala}. Thus, direct
microscopic observations have been used to probe the existence of
aggregates~\cite{haw,varadan}. Confocal microscopy enables the
observation of index-matched suspensions of colloidal particles.
It has been observed that, just after the cessation of flow at
high shear rates, local particle density is extremely
heterogeneous. The highly concentrated regions are supposed to be
responsible for jamming~\cite{varadan,lootens2}. Nevertheless,
confocal microscopy does not allow the observation of rapid motion
of particles. Besides, it is limited to the observation of
index-matched suspensions, and thus modifications of the van
der Waals interactions are required.\\
On the contrary, Diffusing Wave Spectroscopy (DWS), a light
scattering technique adapted to very turbid systems~\cite{weitz},
allows the observation of the dynamics at very short time scales.
 When the sample is illuminated with a coherent light,
multiple scattering occurs. Light propagation can be described by
a random walk, whose transport mean free path, $l^*$, decreases
when the turbidity increases. An interference pattern forms on any
imaginary screen, in particular on the illuminated side of the
cell (backscattering geometry), or on the opposite side
(transmission geometry). The light intensity is spatially
correlated over an area called speckle. Motion of the scatterers
leads to a significant change in the phase of the scattered light,
and hence to a change in the intensity of a single speckle. The
particle dynamics can thus be investigated through the study of
the temporal fluctuations of the scattered intensity on a single
speckle spot. One generally quantify these fluctuations by
computing the intensity autocorrelation function, which is
calculated as a time average. In our case, we are interested in
the dynamics of the particles just after flow cessation. This is a
non-stationary dynamics and thus cannot be studied through a time
average measurement. The multi-speckle diffusive wave spectroscopy
technique (MSDWS) has been introduced to overcome this
limitation~\cite{cipelletti}, as the intensity autocorrelation
function is computed by averaging the intensity fluctuations over
the pixels of a digital camera detector collecting the whole
speckle pattern. Nevertheless, the temporal resolution is limited
by the low frequency of the camera collected images. In our case,
the characteristic time of the system dynamics is too small
to be investigated through this technique.\\
 In this work, we thus use a recently introduced technique, Speckle Visibility Spectroscopy
 (SVS)~\cite{durian} to investigate the
dynamics induced by aggregates formation. The principle of the
measurement is the following~: for a given exposure time, the
faster the dynamics of the suspension, the more the speckle image
is blurred and the less contrasted is the speckle image. Thus the
contrast of an image, computed as the variance of the intensity
distribution across the pixels, allows to explore the dynamics of
the particles~\cite{durian}. The temporal resolution of this
technique is of the order of the exposure duration, much smaller
than the elapsed time between two successive images, and allows
the study of our system under flow. More specifically, we apply a
high shear stress to our suspension, and, after a given time, stop
the stress application. The suspension is continuously illuminated
by a laser beam and by following the contrast of the interference
pattern, we study the particles dynamics under shear and after
shear cessation.

\section{Sample preparation and SVS measurements}
Water suspension of spherical silica particles of diameter $640$
nm, synthesized according to the St\"{o}ber method~\cite{stober},
are studied. To reach more easily the jamming regime the particles
surface is roughened : for a concentrated suspension, it has been
shown that the increased inter-particle contacts and friction
increase the overall viscosity and induce jamming at smaller
stresses than for smooth spheres~\cite{lootens2,tesilootens}. In
order to roughen the particles surface, sodium hydroxide, in a
mass percentage of $19\%$ with respect to the mass of silica, is
added to the suspension, which is then left under stirring for 24
hours. At basic \textit{pH}, the silica depolymerizes
slowly~\cite{iler} and one gets rough particles of the same
diameter. The mean square surface roughness of the particles was
measured by atomic force microscopy and is $6.20\:\:nm$, whereas
it is $0.68\:\:nm$ for the particles before the attack at basic
p\textit{H}~\cite{tesilootens}. The suspension is then rinsed
through centrifugation and redispersed in \textit{Millipore}
water, until \textit{pH} becomes neutral. We then prepared a
suspension of rough particles at a volume fraction of $0.37$. The
sample has an opaque white appearance, so we operate in a multiple
scattering regime.

A stress controlled \textit{Carri-Med} rheometer is used. The cell
is a Couette cell with a rotating internal cylinder of $27.5$ mm
diameter and a fixed external plexiglass cylinder of $30.0$ mm
diameter, which lets the laser beam pass through. A
\textit{Spectra-Physics} Argon polarized laser beam, of wavelength
$\lambda=514$ nm, is expanded and hits the sample with a gaussian
spot size of $6$ mm, at an angle of $\pi/6$ from the normal of the
outer cylinder surface. The light is then multiply scattered by
the suspension and the backscattered light is collected, in a
direction perpendicular to the cell outer surface. The collection
optics consists of a collimating lens that focuses diffused light
onto a diaphragm, which selects a part of it. Finally, a
\textit{Pulnix} CCD camera behind the diaphragm collects the
speckle pattern. The camera device has $768\times 484$ pixels and
can collect the images at a frequency $\nu=15$ Hz. It is
interfaced to a PC provided with a \textit{National Instruments}
card and the data are analyzed in real time using
\textit{LabWindows}. The diaphragm size can be changed in order to
adjust the speckle size and then the ratio of pixels to speckle
areas~\cite{cipelletti}. As the light multiply scattered by the
sample will be depolarized, a polarizer is added between the lens
and the diaphragm in order to minimize direct reflections. All the
measurements were performed in backscattering geometry~; the
dynamics of the particles is thus probed in a volume defined by
the section of the diaphragm ($\simeq 6$ mm $\times 6$ mm) and the
photon penetration depth in the sample. Using a procedure
described elsewhere~\cite{tesilootens}, we measured the photon
mean free path inside the suspension and obtained
$l^*=93\pm\:4\:\:\mu $m. According to DWS theory, the penetration
depth in backscattering geometry is of the order of a few mean
free path~\cite{weitz}. We can thus estimate that the volume
explored in one experiment is of the order of $6$ mm $\times 6$ mm
$\times 0.2$ mm. It thus contains $10^9$ particles, but represents
a small fraction of the entire Couette cell, and in particular,
its depth is approximately one fifth of the gap.

As the particles move, the speckle pattern changes and large
intensity fluctuations occur at each pixel: if the camera exposure
time is long compared to the timescale of speckle fluctuations,
the same average intensity will be recorded for each pixel. On the
contrary, if the exposure time is shorter, the speckle pattern is
visible. This is the principle underlying the SVS technique
\cite{durian} and the key measurable quantity is the variance of
intensity across the pixels. More precisely, we calculate the so
called contrast:
$$C(T)=\frac{\langle I_T^2\rangle_p}{\langle I_T\rangle_p^2}$$
where the $\langle ..\rangle_p$ is an ensemble average over all
the pixels and the intensity $I_T$ is the pixel time-integrated
intensity over the exposure duration $T$. For $T$ much bigger than
the system dynamical timescale the contrast will be 1, in the
opposite limit it will be 2. From another point of view, keeping
$T$ fixed, the contrast will be high if the system dynamics is
slow and low if it's fast with respect to $T$. The exposure
duration of the camera device can vary in the range $64\:\:\mu$s
$-19$ ms. When the exposure duration is varied the laser intensity
is modified in order to keep $\langle I_T\rangle_p$, the
average intensity over the pixels, fixed~\cite{durian2}.\\

\begin{figure}
\begin{center}
\includegraphics[height=6.5 cm]{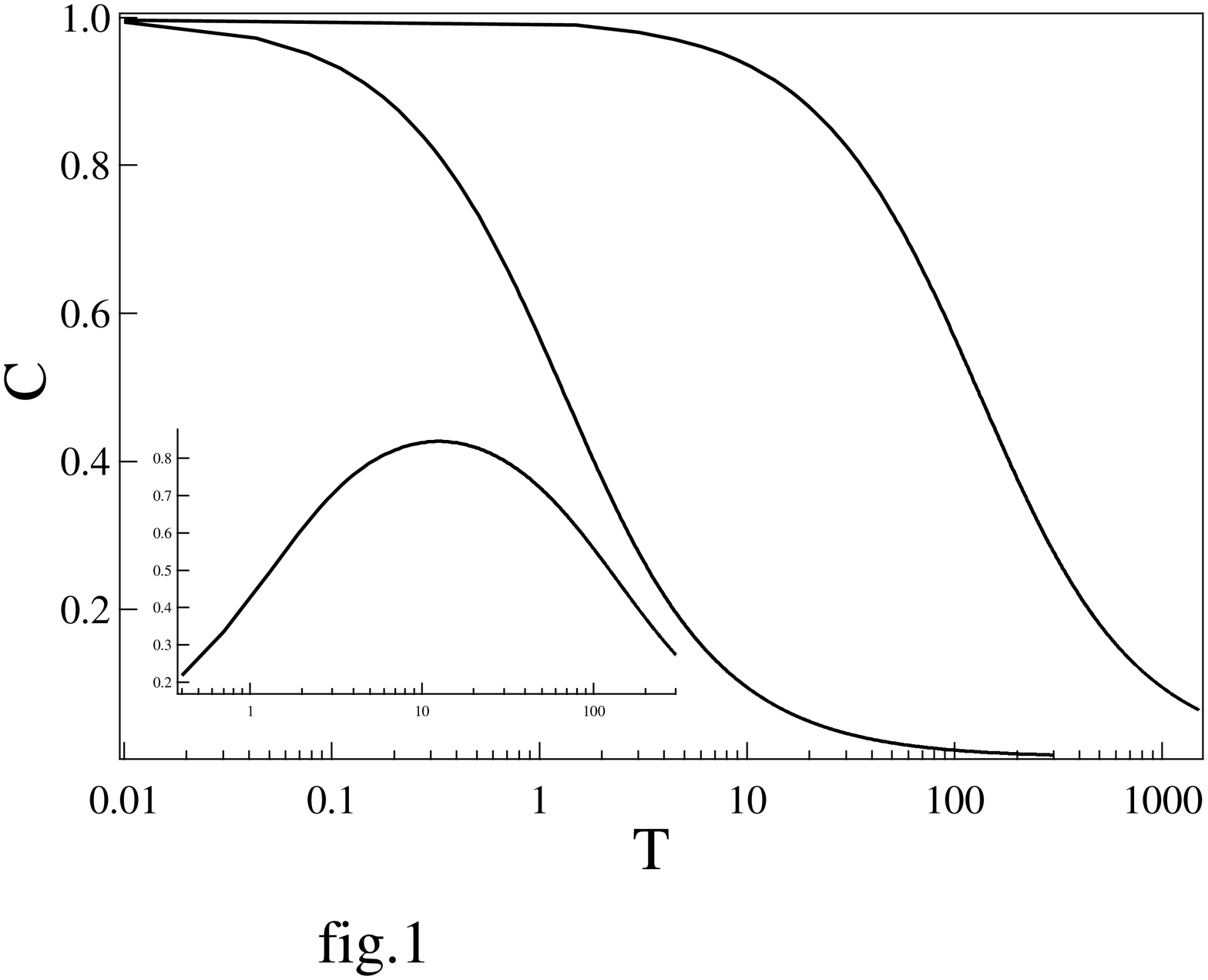}
\caption{Evolution of the contrast with the camera exposure
duration $T$. The dynamics of the electric field correlation
function $g_1$ is assumed to be characterized by a single
relaxation time, $\tau_c$. Continuous curve~:
$\tau_c=\tau_c^0=100$. Dashed curve~: $\tau_c=\tau_c^1=1$.
\textit{Inset~:} difference between the two contrast curves,
$C_{\tau_c^0}(T)-C_{\tau_c^1}(T)$ as a function of $T$. The
maximum difference is obtained for a $T$ value approximately equal
to the geometric average of $\tau_c^0$ and $\tau_c^1$,
$\sqrt{\tau_c^0\tau_c^1}$}\label{C(T)}
\end{center}
\end{figure}

During an experiment, we chose to keep the exposure time constant.
The optimum choice of the exposure time depends on the dynamics of
the observed sample. Let us indeed assume that the dynamics of the
system is characterized by a decorrelation time $\tau_c$ of the
electric field autocorrelation function ($g_1$)~\cite{berne}, and
that $\tau_c$ may take any value between $\tau_c^0$ and
$\tau_c^1$, corresponding to two different values of the contrast,
$C_{\tau_c^0}(T)$ and $C_{\tau_c^1}(T)$. Assuming that $g_1$
exhibits a simple mono-exponential decay, the contrast can easily
be calculated as a function of $T$~\cite{durian}: it is plotted
for the values $\tau_c^0=1$ and $\tau_c^1=100$ in Fig.~\ref{C(T)}.
We wish to observe the maximum variation of the contrast during
the experiment. One needs to find the value of $T$ that maximizes
the difference between $C_{\tau_c^0}(T)$ and $C_{\tau_c^1}(T)$. We
observe that the maximum difference between the two curves is
reached for a time $T=12$, of the order of the geometric mean of
$\tau_c^0$ and $\tau_c^1$, $\sqrt{\tau_c^0\tau_c^1}$. This result
can be easily generalized, as it doesn't change significantly if
we model the decaying electric field correlation function by other
forms (different from a simple exponential), or if we consider a
double decaying correlation function -with one timescale remaining
fixed- to account for another dynamical process present in the
system at a different timescale. We empirically chose for $T$ the
value that maximizes the variation of the contrast during the
experiments, and found that $T=5.08$ ms was the best choice for
our system.

\section{Results}
\begin{figure}
\begin{center}
\includegraphics[height=6.5 cm]{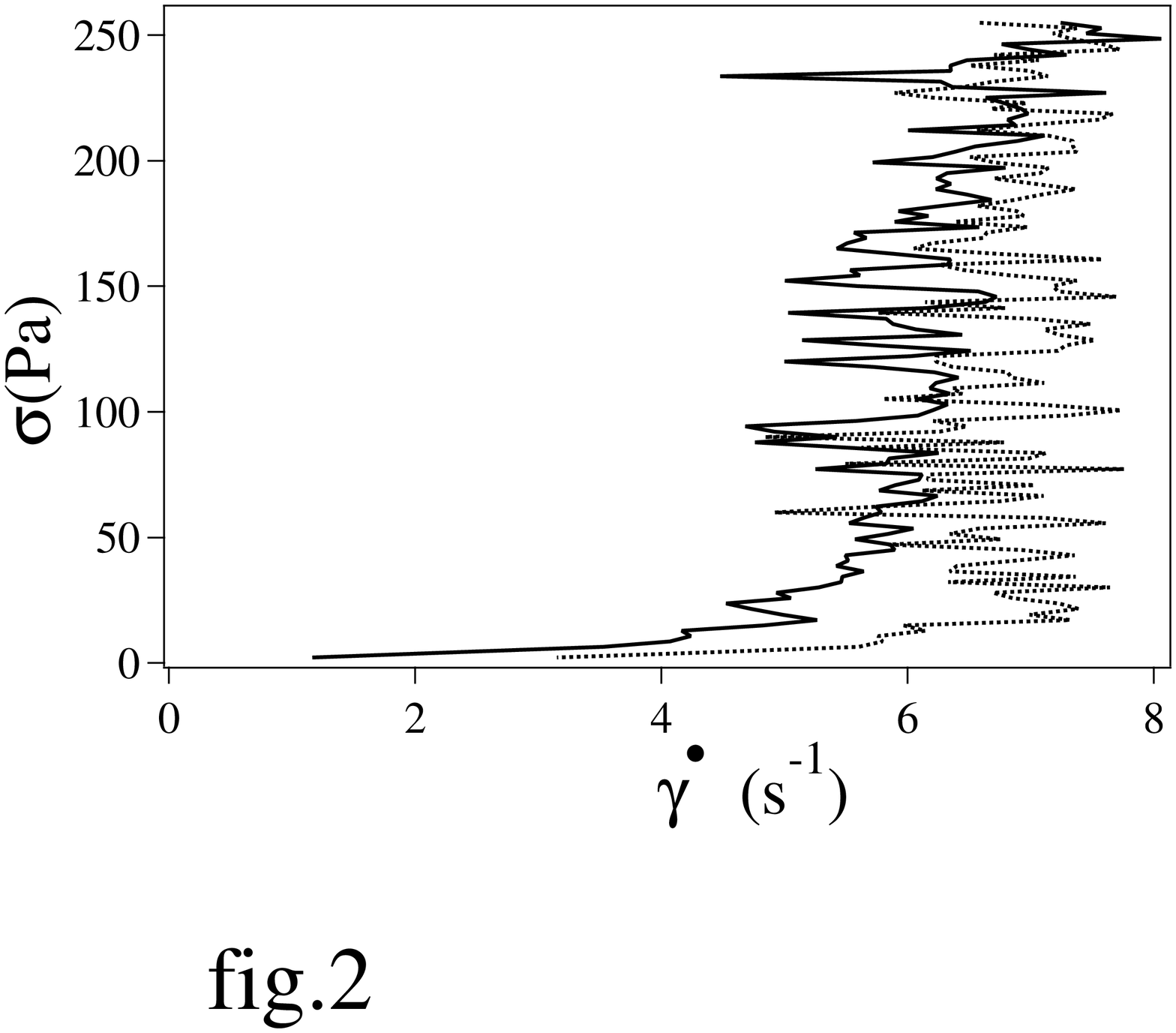}
\caption{Shear thickening of a concentrated suspension of rough
particles. The volume fraction is $\phi=0.37$. Stress is
controlled in Couette geometry, and is increased from $0$ Pa to
$255$ Pa, at a rate of $2.1$ Pa.s$^{-1}$ (continuous curve). Then,
the stress is decreased down to zero with the same absolute rate
(dashed curve). Shear thickening occurs for $\sigma\gtrsim 20$ Pa.
A slight hysteresis is observed when the stress is
decreased.}\label{rheology}
\end{center}
\end{figure}

The mechanical properties of the sample and the occurring of
jamming are illustrated by the rheological measurement reported in
Fig.~\ref{rheology}, in which the sample undergoes a ramp of
stress. For small stresses, the shear rate increases smoothly with
the stress; then, when the stress reaches a critical value, a
transition occurs to a different regime, where the shear rate
starts fluctuating around a fixed value. We will call this the
shear jamming regime. If a decreasing ramp of stress is applied
after the rising ramp, a slight hysteresis in the stress
\textit{vs} shear rate curve is observed.

\begin{figure}
\begin{center}
\includegraphics[height=6.5 cm]{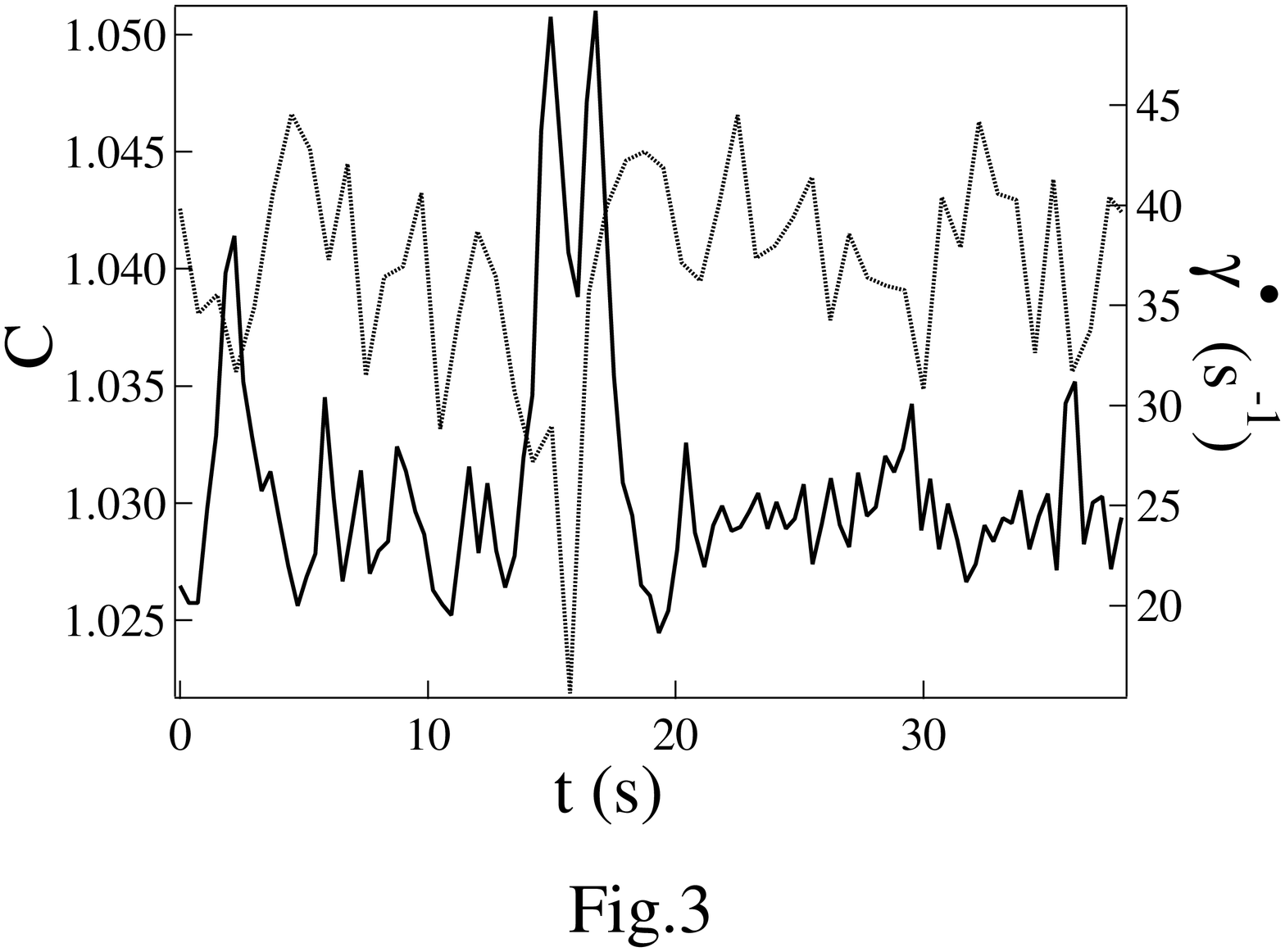}
\caption{Fluctuations of the gradient (dotted line) and the
contrast (solid line) under an applied stress of $50$ Pa. Low
values of the gradient are associated to high values of the
contrast.}\label{fluctuations}
\end{center}
\end{figure}

Let us now apply a constant stress of $50$ Pa, above the jamming
transition (Fig.~\ref{fluctuations}). First, we observe that the
shear rate is not stationary and exhibits huge
fluctuations~\cite{lootens}. The observed contrast also fluctuates
and its fluctuations are correlated to the shear rate
fluctuations. When the sample is under shear, the faster and
dominant movement of the particles is due to the flow. Assuming a
linear shear rate, the typical timescale characterizing the
scattered light intensity fluctuations is
$\tau_s=\sqrt{10}/(\textbf{k}_0 l^*)1/\dot{\gamma}$, where
$\textbf{k}_0=2\pi/\lambda$ is the laser beam wave vector,
$\dot{\gamma}$ is the shear rate and $l^*$ is the photon transport
mean free path in the medium~\cite{wu}. So, when the flow velocity
decreases due to the shear rate fluctuations in the jamming
regime, as the particles move slower, the contrast will be higher
and viceversa.

\begin{figure}
\begin{center}
\includegraphics[height=6.5 cm]{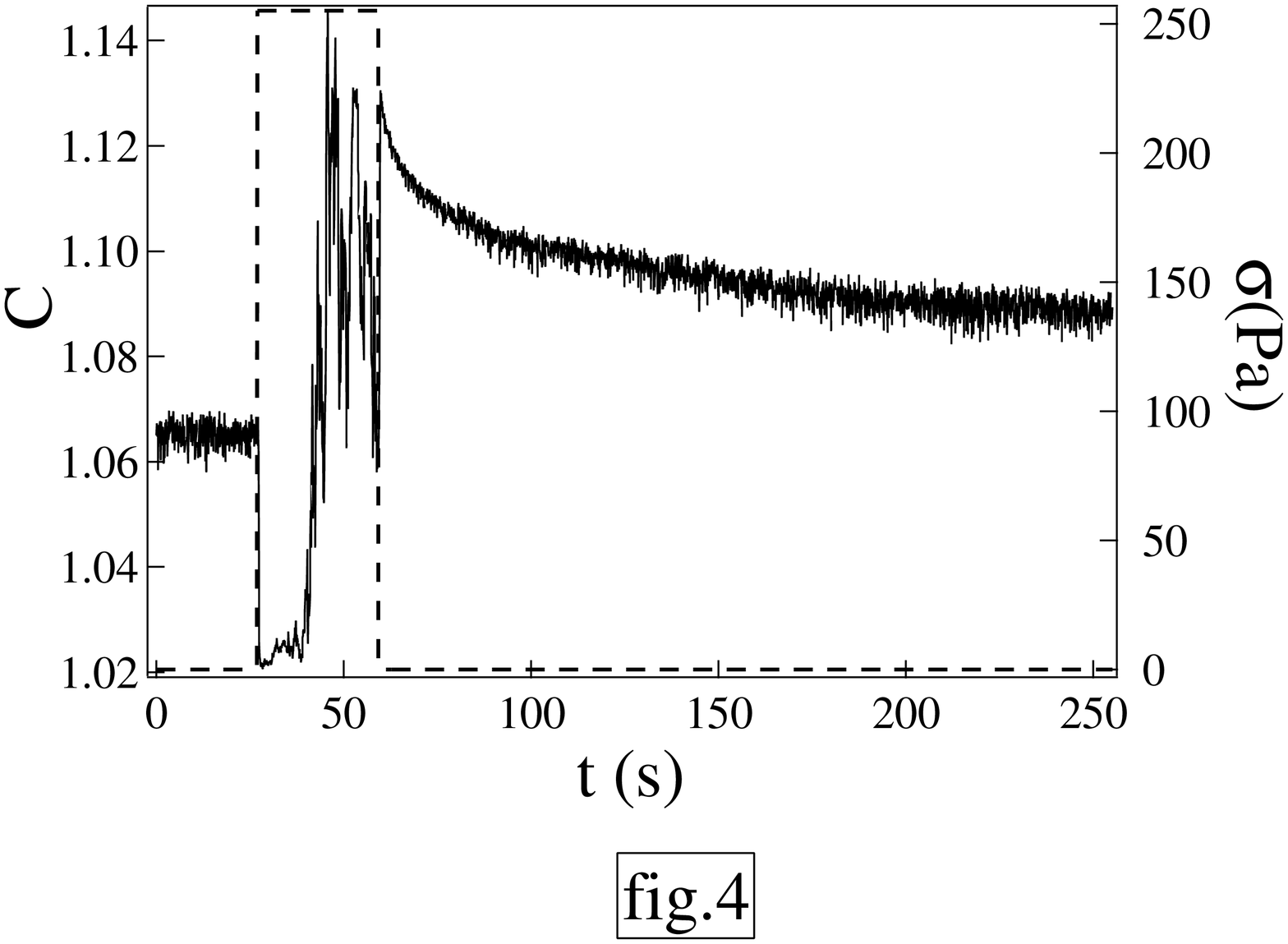}
\caption{Contrast behavior before, during and after the
application of a stress $\sigma=255$ Pa during $30$ s. The stress
history is plotted in dashed line. During the stress application,
strong contrast fluctuations are observed. After the shear
cessation, the contrast exhibits an overshoot and then relaxes to
a constant value.}\label{overshoot}
\end{center}
\end{figure}

Under high stress values, the dynamics of the particles under
shear may even be slower than the dynamics at rest. In
Fig.~\ref{overshoot}, we report an experiment in which a high
stress value in the jamming regime, $\sigma=255$ Pa, is applied
for $30$ s; then the application of stress is stopped. Before the
shear, the contrast at rest is equal to $1.06$ and its relative
fluctuations, calculated as the standard deviation of the signal
divided by its average value diminished by $1$, are $4\%$. As soon
as the stress is applied, the contrast drops to a smaller value
$1.02$; then, the suspension jams and huge contrast fluctuations
occur. Remarkably, these fluctuations lead to contrast values
higher than the contrast value at rest before or after the stress
application. This means that during the shear, the particles
velocity becomes sometimes so small that $\tau_s$ is not the
dominant timescale anymore, so another dynamical timescale of the
system can be revealed. Moreover, the dynamics characterized by
this sometimes emerging timescale is slower than the dynamics of
the system at rest. The particles thus organized themselves under
flow in such a way that their motion is slower than their free
motion at rest. This is an important result as it is a direct
evidence of the formation of a flow-induced structure whose
dynamics is slower than the dynamics of the suspension's at rest.

The dynamical properties of particles aggregates and their
relaxation time are studied. We thus follow the evolution of the
contrast after cessation of stress in the jamming regime. As soon
as the stress application is stopped, we observe
(Fig.~\ref{overshoot}) an overshoot of the contrast: just after
stress cessation, the contrast value is much higher than the one
at rest. It then slowly decreases to a constant value, which is
not necessarily the same value it had before the shear. We observe
that the contrast signal is very noisy. Let us moreover consider
the noise amplitude along a relaxation curve
(Fig.~\ref{overshoot})~: just after shear cessation ($60<t<80$ s),
the amplitude of the fluctuations are smaller than after complete
relaxation ($t>200$ s). The origin of the noise is discussed
below. When the same measurement is repeated under the same
conditions, the relaxation curve of the contrast may exhibit
different properties (Fig. \ref{relaxation}): its amplitude, its
noise, the final value and the characteristic relaxation time vary
with the measure; while sometimes the overshoot cannot be revealed
either, as one of the three curves shows.

\begin{figure}
\begin{center}
\includegraphics[height=6.5 cm]{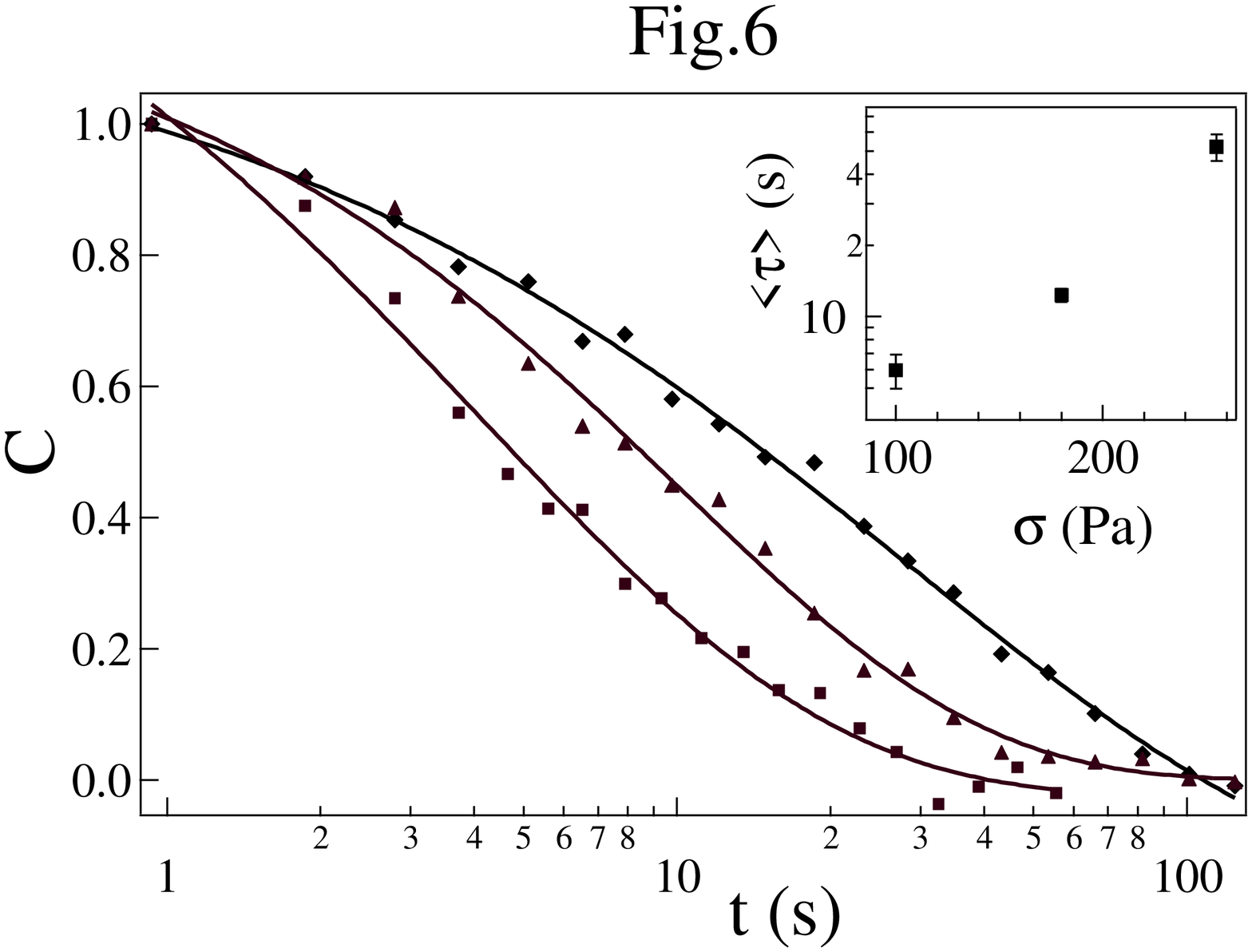}
\caption{Normalized and averaged contrast relaxations curves after
the application of a constant stress $\sigma$ during $30$ s. Three
values of the applied stress are studied.
$\blacksquare$~:$\sigma=100$ Pa, $\blacktriangle$~:$\sigma=180$ Pa
and $\blacklozenge$~:$\sigma=255$ Pa. Each curve is fitted with a
stretched exponential form~: $C(t)=e^{-(\frac{t}{\tau})^\beta}$.
\textit{Inset~:} average relaxation time,
$\bar{\tau}=\int_0^{\infty}exp[-(t/\tau)^\beta]\:dt=\tau/\beta
\:\:\Gamma(1/\beta)$ as a function of the applied stress.
}\label{3stress}
\end{center}
\end{figure}

Though the contrast signal is very noisy, we investigated
qualitatively the dependence of the relaxation contrast behavior
on the applied stress value in the jamming regime. The evolution
of the contrast after the shear stop has been studied for three
different applied stress over the critical stress and for each
stress a set of $8$ measurements under the same conditions has
been performed. The sample is first sheared under a constant
stress $\sigma\in \{100,\: 180,\: 255\}$ Pa for $30$ s, then
stress application is ceased; the contrast value is recorded
during the entire experiment. The curves which were not showing an
overshoot have been dropped. We used the following criterium~: we
selected only the curves whose noise was smaller than the
half-amplitude of the contrast decrease. The noise is measured as
the standard deviation of the contrast when it reaches the final
plateau value, \textit{i.e.} when it is the highest. For the
larger stress value, $\sigma=255$ Pa, none of the $8$ curves had
to be dropped, for $\sigma=180$ Pa, $2$ of the set were dropped
and $4$ for $\sigma=100$ Pa. In order to compare the relaxation
times after the application of different stresses, we averaged the
set of remaining contrast data for each stress value. As the
contrast relaxation has an exponential behavior, we applied a
logarithmic binning procedure to each averaged curve, in order to
reduce noise at long time. Firstly, the points of the curve have
been averaged in groups of ten, then they are further averaged in
order to obtain a curve with $25$ points logarithmically spaced on
the x-axis. The obtained curves of the averaged contrast evolution
\textit{vs} time after shear cessation for the three different
stress values are plotted in Fig.~\ref{3stress}, where, to be
better compared, they have been normalized between $0$ and $1$.
The curves are then fitted by a stretched exponential
$C(t)=\exp(-t/\tau)^\beta$. The value of the $\tau$ parameter
varies between $4$ s for the smaller stress and $29$ s for the
bigger stress. As the value of the $\beta$ parameter does not
remain constant, but varies between $0.5$ and $0.7$, the average
relaxation time $\langle\tau\rangle$ is calculated~:
\begin{equation}
\langle\tau\rangle=\int_0^{\infty}\textrm{e}^{-(t/\tau)^\beta}\:dt=\frac{\tau}{\beta}
\:\:\Gamma(\frac{1}{\beta})
\end{equation}\noindent
where $\Gamma$ is the Euler gamma function. The values of
$\langle\tau\rangle$ are plotted as a function of the stress in
Fig.~\ref{3stress}~\textit{inset}. Thus, at high stresses, where
particles  aggregates formation has been demonstrated, the
contrast relaxation after shear cessation gets slower for
increasing applied stress.
\section{Discussion}
These experiments thus lead to the following observations~:
\begin{itemize}
    \item the contrast fluctuates under shear, and may reach
    values higher than values at rest (Fig.~\ref{overshoot}),
    \item contrast relaxation curves after shear cessation exhibit huge noise amplitude (Fig.~\ref{relaxation}),
    \item the contrast relaxation time after shear cessation is an increasing function
    of the applied stress (Fig.~\ref{3stress}).
\end{itemize}

\begin{figure}
\begin{center}
\includegraphics[height=6.5 cm]{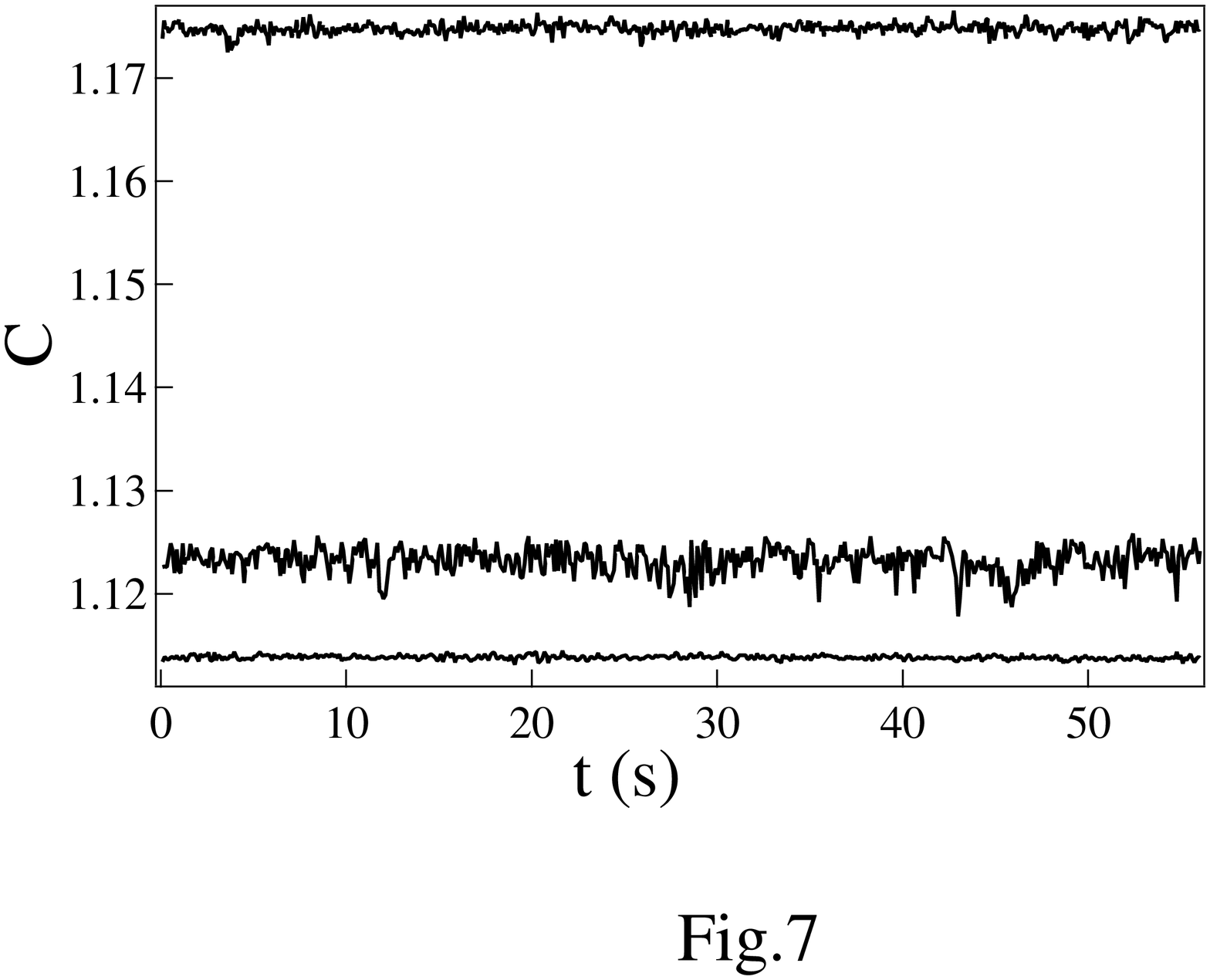}
\caption{Contrast evolution for a system at rest, for different
samples and different camera exposure times $T$. From top to
bottom: silica spheres suspension for $T=191\:\:\mu s$, silica
spheres suspension for $T=19.1\:\:ms$ and water solution of latex
for $T=19.1\:\:ms$.}\label{noise}
\end{center}
\end{figure}

When put together, these observations give insight into the
aggregates formation and disruption. But, first of all, let us
discuss the origin of the high noise of the contrast signal. It is
a consequence of both the properties of the system, and of our
choice of the shutter duration $T=5.08$ ms, that maximizes the
amplitude of the contrast relaxation after shear cessation.
Indeed,  the contrast is very noisy if compared to the amplitude
of the relaxation (Fig. \ref{relaxation}); this prevents us from
making systematic measurements with an easy quantitative analysis
of the results. This noise is not due to the setup, but is an
intrinsic characteristic of the sample. This is shown in
Fig.~\ref{noise}, where we plotted the contrast, measured at the
camera exposure duration $T=19.1$ ms, for two samples at rest: the
suspension of silica particles used in our work and a water
solution of latex, taken as reference sample. The relative noise
was calculated as the ratio between the standard deviation of the
signal and the difference between its average value and $1$. We
obtained a value of $0.011$ for the noise of the silica sample and
of $0.0035$ for the one of the reference sample, which is then
nearly an order of magnitude smaller. Besides, for a given sample,
the noise of the contrast signal depends on the chosen $T$. The
amplitude of the noise decreases when the exposure time $T$
decreases. Thus, the relative contrast noise for a silica
suspension for $T=191\ \mu$s is smaller by a factor of $4$ than
the noise for $T=19.1$ ms (Fig.~\ref{noise}). Nevertheless, as
explained above, the value of $T$ was chosen in order to maximize
the contrast variation during the entire experiment. For that
value ($T=5.08$ ms), the amplitude of the noise is similar to that
measured at $T=19.1$ ms.

\subsection{Aggregates formation under flow} When the suspension is submitted to a constant stress,
high values of the contrast are associated to low values of the
shear rate (Fig.~\ref{C(T)}). The contrast increase events are
thus due to the formation of aggregates of macroscopic size that
hinder the flow, and induce a drop of the shear rate. Moreover,
the observed contrast under shear may reach higher values than
contrast values at rest (Fig.~\ref{overshoot}). This observation
has two consequences~: dynamics of the particles inside the
aggregates is slower than their dynamics at rest, and the volume
occupied by particles belonging to aggregates is large enough so
that position fluctuations of free particles inside the
illuminated volume do not blur the speckle image.
\subsection{Spatial heterogeneities} 
\begin{figure}
\begin{center}
\includegraphics[height=6.5 cm]{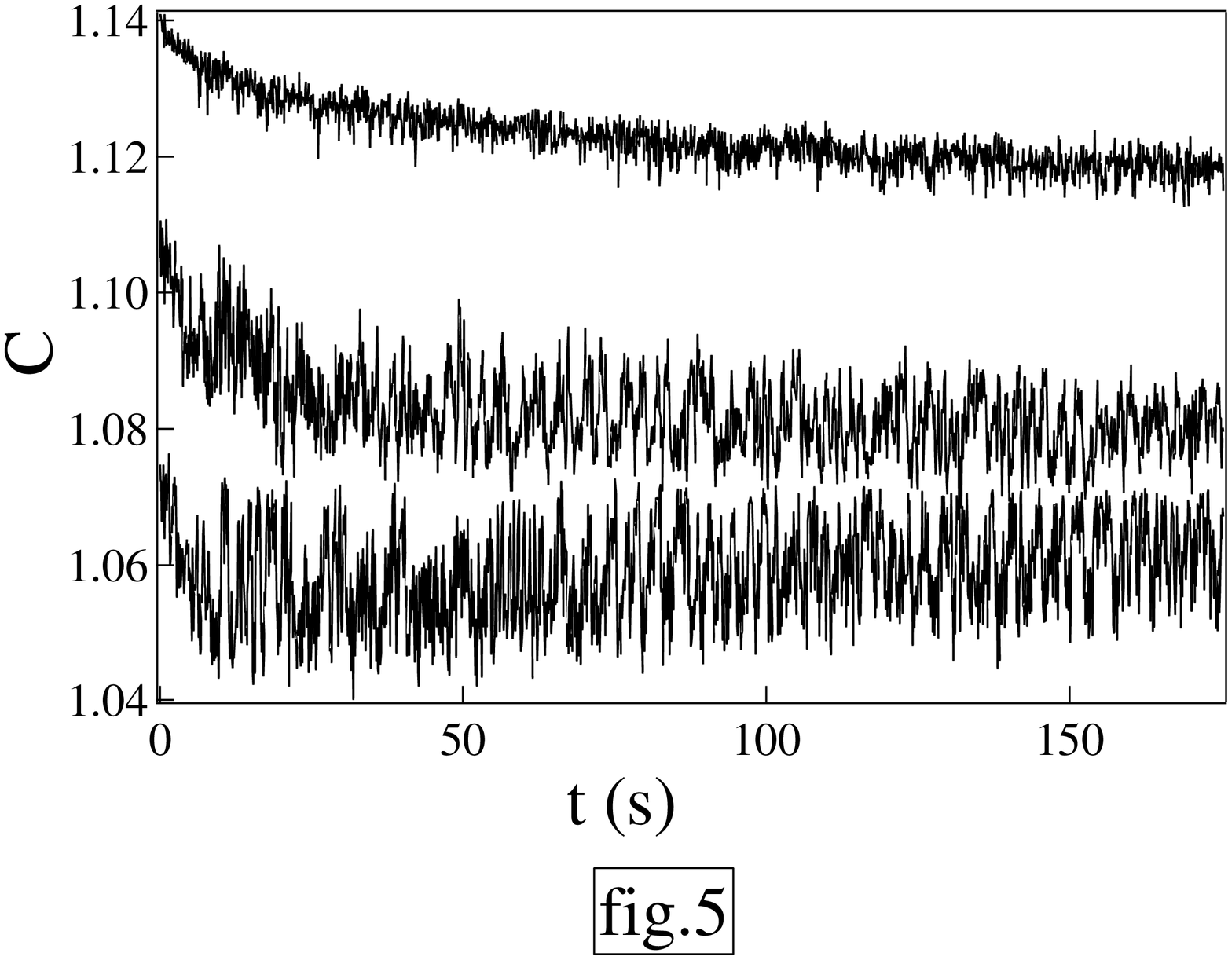}
\caption{Contrast relaxation, soon after the application of a
stress $\sigma=180$ Pa during $30$ s, for three different
measurements performed under the same conditions. The measurements
are highly non-reproducible. The overshoot of the contrast is not
always observed, as exemplified by the bottom curve. The two other
curves exhibit an overshoot, but its amplitude, its characteristic
decorrelation time, the value of the baseline and the noise of the
contrast signal vary with the measurement.}\label{relaxation}
\end{center}
\end{figure}

In Fig.~\ref{overshoot} we observed that, once the
stress application is stopped, the contrast plateau value after
the relaxation is different from the contrast value at rest before
stress application. Besides, Fig.~\ref{relaxation} shows that this
plateau value varies at each measurement. As the ensemble of
measurements presented in Fig.~\ref{relaxation} are taken with the
same sample under the same conditions, it can be considered as a
sampling in which a different region of the system after shear
cessation is observed during each measurement. The presence of
spatial heterogeneities in the system, consisting in slower and
faster regions, due to the very high concentration of the
suspension may be responsible for this lack of reproducibility.
Indeed, if the heterogeneities length-scale is of the order of the
illuminated volume size, the non reproducibility of the contrast
value at rest may be easily explained. After each shear cessation,
the illuminated region would be characterized by a different local
concentration and then a different dynamical timescale inducing
different contrast values. Fig.~\ref{relaxation} thus gives us
some qualitative insight into the aggregates size. The amplitude
of the relaxation of the contrast depends on the relative size of
the aggregates in the illuminated volume. It has been shown by
numerical simulations~\cite{loan}, that the presence of aggregates
spanning the entire system is characteristic of the jamming
regime. The varying amplitude of the contrast relaxation for
different measurements implies that the aggregates size is
comparable to the illuminated region, thus confirming the
mechanical results. Unfortunately, a study of the aggregate size
distribution is not possible, due to the fact that the illuminated
region, representing a fraction of the entire cell gap, may
contain only a part of an aggregate. Moreover, for the lowest
studied values of the applied stress, the contrast doesn't always
exhibit an overshoot after the application of shear. There must
thus exist some regions, at least as large as the illuminated
volume, without any of the macroscopic aggregates responsible for
the jamming. This gives a lower limit of the average distance
among the aggregates, which must be at least of the maximum linear
size of the illuminated region, \textit{i.e.} $6$ mm.
\subsection{Contrast fluctuations} The existence of such heterogeneities allows us to
understand the observed large fluctuations of the contrast and the
evolution of their amplitude during relaxation. The decorrelation
time $\tau_c$ for $g_1$, characterizing the dynamics of the
system, will have a certain distribution $\Delta\tau_c$.
Remembering that the contrast can be considered as a function of
$g_1$~\cite{durian}, a large decorrelation time distribution will
induce large fluctuations in the contrast signal. Moreover, the
amplitude of this fluctuations will also depend on the value of
$T$ relatively to the decorrelation time. Thus, let us assume that
the system is characterized by a decorrelation time belonging to
$[\tau_0-\Delta\tau_0,\tau_0+\Delta\tau_0]$ with
$\Delta\tau_0=\tau_0/10$ soon after the shear stop, and by a
characteristic time $\tau_1>\tau_0$ with distribution
$\Delta\tau_1=\tau_1/10$ when it has relaxed to equilibrium. In
order to obtain the maximum value of the contrast, we chose
$T=\sqrt{\tau_0\tau_1}$, as detailed in Fig.~\ref{C(T)}. Choosing
for instance $\tau_0=1$ and $\tau_1=100$, the amplitude of the
contrast fluctuations may be computed. One gets that fluctuations
for dynamics faster than the exposure time are $\sim 15$ times
larger than fluctuations for slower dynamics. This computation
agrees with the contrast fluctuations observed in
Fig.~\ref{overshoot}, where the contrast noise soon after the
shear cessation ($60<t<80$ s), \textit{i.e} for the slowest
dynamics, is smaller than the noise after the contrast relaxation
($t>200$ s), when the dynamics is the fastest.
\subsection{Stress dependency}
Finally, from the study of the contrast relaxation at different
stresses, we observed that, the higher the stress, the more
frequently the contrast overshoot shows up. Thus, the volume
fraction of macroscopic aggregates in the system increases with
the applied stress, leading to an increase of the probability to
observe one of them when stress is ceased. Nevertheless, we did
not observe any dependency of the amplitude of the contrast
relaxation after shear cessation on the applied stress. As long as
the aggregates are smaller than the illuminated volume, this
amplitude is a measure of the volume fraction of the aggregates
inside the illuminated volume. As a consequence, the observed
increase of the aggregates volume fraction would not be due to an
increase of the aggregates size, but of the aggregates number.
Nevertheless, direct confocal microscopy observations suggest that
the jammed regions may be larger than the diameter of our beam
($6$ mm)~\cite{lootens2}. Thus, an increase of the aggregates size
as a function of the applied stress cannot be
ruled out.\\
Two phenomena may thus be responsible for our last result. We
observed that, on average, for larger applied stresses, the
relaxation time of the aggregates is longer
(Fig.~\ref{3stress}~\textit{inset}). On one hand, an increase of
the size of the aggregates, as the applied stress value is
increased, may occur. On the other hand, colloids organization and
volume fraction inside the aggregates may depend on the applied
stress. Thus, numerical simulations of concentrated colloids under
shear~\cite{melrose1} show that the distance between neighboring
particles inside the aggregates decreases when the applied stress
increases. If similar behavior occurs in our system, as also
indicated by confocal microscopy observations~\cite{lootens2}, the
volume fraction of particles inside aggregates increases with the
applied stress, which favors a slowing down of aggregates
disruption.

\section{Conclusion}
We investigated the aggregates relaxation after shearing a
suspension of concentrated silica particles in the jamming regime.
Under flow, the observed peaks in the contrast showed that
aggregates form. The dynamics of the particles under flow may even
be slower than their dynamics at rest. When the shear is stopped,
the contrast relaxes back to a lower value, giving evidence of the
reversibility of the aggregates. The suspension under flow and
just after flow cessation is extremely heterogeneous. The
characteristic size of the heterogeneities is of the order of
magnitude of the illuminated volume in the SVS experiment,
\textit{i.e.} of the order of the cell thickness. The contrast
signal thus exhibits huge fluctuations and measurements are not
easily reproducible, as they depend on the particular volume
sampled during each experiment. Nevertheless, by carefully
averaging the contrast relaxation over several measurements, it
was observed that the higher the applied stress, the slower the
aggregates relaxation. Several phenomena may be responsible for
this behavior~: an increase of the aggregates size and volume
fraction, as also indicated by the increase of the occurrence of
contrast overshoot with the applied stress, or a decrease of
intercolloidal distances inside aggregates, as suggested by
numerical simulations~\cite{melrose1}. Further experiments should
allow to choose among the interpretations proposed.

The authors wish to thank R. Di Leonardo and G. Ruocco for
fruitful discussions.

 
\end{document}